\documentclass[pre,twocolumn,showpacs]{revtex4}

\usepackage{graphicx}

\begin{document}

\title[Version 8]{Interaction of Ising-Bloch fronts with Dirichlet boundaries}

\author{A. Yadav}
 \email{yadav@phys.lsu.edu}
\author{D. A. Browne}
\affiliation{Department of Physics and Astronomy,
Louisiana State University,
Baton Rouge, LA 70803--4001}


\begin{abstract}
We study the Ising-Bloch bifurcation in two systems, the Complex Ginzburg
Landau equation (CGLE) and a FitzHugh Nagumo (FN) model in the presence of
spatial inhomogeneity introduced by Dirichlet boundary conditions. It is
seen that the interaction of fronts with boundaries is similar in both
systems, establishing the generality of the Ising-Bloch bifurcation. 
We derive reduced dynamical equations for the FN model that explain 
front dynamics close to the boundary. We find that front dynamics 
in a highly non-adiabatic (slow front) limit is controlled by fixed points
of the reduced dynamical equations, that occur close to the boundary.
\end{abstract}

\maketitle

\section{Introduction}

In spatially extended reaction--diffusion systems far from equilibrium,
the interplay of the diffusion and reaction processes is frequently
associated with the formation of spatial or temporal patterns
in the concentration fields \cite{lee,lee2,lee3,meron1,haas,Li}.
One such example is a front-like structure connecting two different
homogeneous steady states.  In a bistable system, where both steady states
are stable against small perturbations, these fronts can undergo
bifurcations, known as nonequilibrium Ising--Bloch bifurcations,
where a stationary Ising front exchanges stability with a pair
of counterpropagating Bloch fronts.
This bifurcation has been observed in several 
chemical reactions\cite{meron1,haas,Li} and also in liquid crystals
\cite{Coullet,Frisch,kai} subject to an external time--dependent
perturbation.

Two models of this Ising--Bloch bifurcation have been 
extensively studied in this context.
One is the
parametrically forced Complex Ginzburg Landau equation (CGLE)  \cite{Coullet}.
This system describes nematic liquid crystals subjected to a rotating
magnetic field and a high frequency electrical field \cite{Frisch}.
 The CGLE is often used 
in spatially extended systems to describe the dynamics close to an
 oscillatory instability (Hopf bifurcation). The other is a FitzHugh Nagumo
 (FN) model \cite{meron1,meron2,meron3,meron4,meron5},
which qualitatively models various chemical
reactions\cite{Dulos,Ouyang,epstein}. Front solutions in this model
have been extensively investigated, specially when translational
invariance is broken by the presence of spatial inhomogeneities,
 which is often the case in realistic experimental
situations. This includes the ways in which
one Bloch front can be perturbatively changed to the other (leading
to front reversal) \cite{meron2,meron3,meron4,meron5}.
Particularly, one such scenario for front reversals and other
exotic nonuniform (variable velocity) front
motion like breathing, involves the breaking of translational
invariance by zero flux boundary conditions \cite{meron1}. 
This nonuniform motion of fronts is explained by the presence of 
uniform velocity front solutions (nullclines of the FN model partial
differential equations) to which faster or slower moving fronts relax
adiabatically. 

In this paper we examine nonuniform front motion in the case when translational
invariance is broken by imposing
fixed chemical concentrations at the boundary of a reactor (Dirichlet
boundary conditions). An adiabatic description of nonuniform front
motion along the lines of \cite{meron1} is inadequate. It relies
on nullclines and fails to explain the jumps from one nullcline to another or
the influence of fixed points. Therefore, we employ a dynamical approach
which satisfactorily accounts for jumps and fixed point influence.
We also establish the generic nature of nonuniform front motion in
the presence of spatial inhomogeneities by studying 
it in two distinct systems, the FN and CGL models.

 The interaction of traveling fronts with boundaries for Dirichlet
boundary conditions shows several new features.
We see a transition from front reversal to trapping of an incoming Bloch
front on its approach to a boundary as a function of boundary values.  
We also find that trapped fronts and reversed fronts can coexist for 
certain boundary conditions. Finally, we derive reduced dynamical equations
that explain the features mentioned above.

In section II of this paper we review the relevant details of the
CGL and FN models. Section III presents our numerical study
of the two models. Section IV contains an analytic study of 
front interactions with boundaries for the FN model. Our conclusions 
are listed
in Section V.

\section{The Models}

The parametrically forced CGLE is the generic model describing the
slow phase and amplitude modulations of a spatially distributed assembly
of coupled oscillators near its Hopf bifurcation \cite{vansarloos}. This
assembly of auto-oscillators is parametrically forced at twice its 
natural frequency and can be written as,
\begin{eqnarray}
{\partial A\over \partial t}& = &
(\mu+i\nu)A + (1+i\alpha)\nabla^2 A\nonumber\\
& &
- (1+i\beta)|A|^2 A + \gamma A^{*} + \kappa~.
\label{eq:cgle}
\end{eqnarray}
The complex field $A$ contains the amplitude and phase of the coupled
oscillators, $\mu$ measures the distance from the oscillatory instability
threshold, $\nu$ is the detuning of the forcing term frequency from exactly
twice the Hopf frequency, $\alpha$ and $\beta$ are real
control parameters, and $\gamma>0$ is the forcing amplitude at twice the
natural frequency. 
The right hand side of Eq.~(\ref{eq:cgle}) cannot be written in a variational
form if $\nu$, $\alpha$, and $\beta$ are nonzero. 

The parameter $\kappa$ represents parametric
forcing at the natural frequency of the system. If $\kappa=0$,
Eq.~(\ref{eq:cgle}) has a the parity symmetry $A\to -A$, 
and the nonequilibrium Ising--Bloch bifurcation 
is a symmetric pitchfork \cite{Coullet,meron7} with the front velocity as the 
order parameter. Zero velocity Ising walls lose stability to counterpropagating
Bloch walls, as the bifurcation parameter $\gamma$ crosses its critical value.
The pitchfork unfolds into a saddle node bifurcation for a nonzero $\kappa$ 
where along with the stable Ising wall, a stable and unstable pair of Bloch 
walls appear at the bifurcation.

The CGLE Eq.~(\ref{eq:cgle}) has trivial homogeneous solutions
$(A(x)=\pm \sqrt{\mu+\gamma} + 0i)$ in the variational case $(\alpha=\beta=
\nu=0)$ making it bistable. Solutions connecting these trivial states are,
\[
A(x)=\sqrt{\mu+\gamma}\tanh\left(\sqrt{\frac{\mu+\gamma}{2}}x\right)+0i,
\] which is the Ising wall solution characterized by a zero imaginary
part, and

\[
A(x)=\sqrt{\mu+\gamma}\tanh(\sqrt{2\gamma}x) +
i\sigma\sqrt{\mu-3\gamma}\mbox{sech}(\sqrt{2\gamma}x),
\] the two Bloch wall solutions distinguished by their 
respective chirality $ \sigma=\pm 1$.
In the case when nonvariational parameters $(\alpha, \beta,\nu)$ 
are non-zero, and $\gamma^2>{(\nu-\beta\mu)}^2/(1+\beta^2)$, 
Eq.~(\ref{eq:cgle}) is still bistable and has trivial solutions,
\begin{eqnarray}
A & = & R\exp(i\phi)\nonumber\\
R^2 & = & \frac{\mu+\beta \nu +[\gamma^2(1+\beta^2)-(\nu-\beta
\mu)^2]^{1/2}}{(1+\beta^2)}\nonumber\\
\cos(2\phi) & = & (-\mu+R^2)/\gamma,\nonumber\\
 \sin(2\phi)& = &(\nu-\beta R^2)/\gamma.
\label{eq:trivcgle}
\end{eqnarray}
Nonvariational Ising and Bloch wall solutions are qualitatively different
from their variational counterparts. Nonvariational Bloch walls move as a
result of the breaking of chiral symmetry \cite{Coullet}, whereas variational
Bloch walls are stationary. Chirality breaking is unlike other
mechanisms of front motion, where a globally stable state 
invades an unstable or metastable state \cite{saa1,saa2}. In the perturbative
 limit when $(\alpha, \beta,\nu)$ are small, the 
velocity of Bloch fronts can be written as
\begin{eqnarray}
c=\chi \left[{\mu+\gamma}\over {2\gamma}\right]^{1 \over 2}
\left.{3\pi}\over {2(3\mu-\gamma)}\right.
[-\nu+\beta\mu+(\alpha-\beta)\gamma]~,
\label{eq:cglvel}
\end{eqnarray}
where $\chi=\pm \sqrt{3(\gamma_c-\gamma)}$~. The Bloch wall velocity
$c$ is proportional to $ \sqrt{\gamma_c-\gamma}$ as expected for 
a pitchfork bifurcation.

Now we look at the FN model, which also shows a front bifurcation. This
is a simple two component model and has been thoroughly analysed in 
Ref.~\cite{meron2,meron3,meron4,meron5,meron8} in the context of 
this bifurcation. It has been widely used to model patterns
in reactions like the Belousov-Zhabotinsky (BZ)
reaction \cite{kuhn1,kuhn2,nagy1,nagy2,woods}, Ferrocyanide-Iodate-Sulfite
(FIS) reaction \cite{meron1} and
Chlorite-Iodide-Malonic-Acid reaction(CIMA)
\cite{Dulos,Ouyang,epstein}. The two component reaction-diffusion system,
with $v(x,t)$ impeding the production of $u(x,t)$, is given
by
\begin{eqnarray}
{\partial u\over \partial t}&=&\epsilon^{-1}(u -u^3 -v) +
\delta^{-1} u_{xx}\nonumber\\
{\partial v\over \partial t}&=& (u-a_1v-a_o) +  v_{xx}.
\label{eq:fitz}
\end{eqnarray}
The parameters $\epsilon$ and $\delta$ differentiate the time scales and 
space scales of the two fields respectively. The parameters $a_1$ and
 $a_o$ characterize 
renormalized local reaction parameters, possibly after an adiabatic elimination
of faster reacting species. Equations ~(\ref{eq:fitz}) are in
general nonvariational except for certain specific parameter values.
The parameter $a_o$ is analogous to
the parameter $\kappa$ in the CGL equation, and it controls the 
symmetry of the front bifurcation.
 If $a_o= 0$, the FN model
undergoes a symmetric front bifurcation represented by a pitchfork. For
a nonzero $a_o$ the pitchfork unfolds into a saddle node as in the CGLE. 
A notable difference between the two systems is the
presence of the parameter $\delta$ in the FN model. This
parameter affects the relative spatial extent of the fronts connecting
the trivial homogeneous solutions of Eq.~(\ref{eq:fitz}). 
Thus, by choosing a suitable $\epsilon/\delta$ ratio, the connecting 
fronts of one
of the fields can be made very sharp compared to the other. This is not
 possible in the CGLE, where fronts for both the real and imaginary 
parts have the same spatial extent.

Trivial homogeneous solutions to the FN model are
\begin{equation}
u_c=\pm\sqrt{\frac{a_1-1}{a_1}}\quad v_c=\frac{u_c}{a_1}~.
\label{eq:trivfitz}
\end{equation}
Similar to the CGLE, the FN model has Ising and Bloch fronts as its solutions,
that bifurcate in a pitchfork. The bifurcation parameter in the FN model
may be chosen to be $\eta=\sqrt{\epsilon\delta}$. The critical
value of this parameter is $\eta_c=(3/\sqrt(2)2){(a_1+1/2)}^{3/2}$ \cite{meron4}
. The Bloch
wall velocities given by
\begin{equation}
c^2=\frac{6\sqrt{2}}{\eta_c^{2}(\sqrt{a_1+1/2})}(\eta_c-\eta)~, 
\label{eq:fitzvel}
\end{equation}
are proportional to $\sqrt{\eta_c-\eta}$, the deviation 
of the bifurcation parameter from its critical value, as  expected
for a pitchfork bifurcation.

As discussed above, both these models have common features associated
with the front bifurcation.
This forms the basis of their comparative study in the forthcoming sections.
Front dynamics in both these systems can be represented by the system of 
equations,
\begin{eqnarray}
\dot{x}&=&c\nonumber\\
\dot{c}&=&(\rho_c-\rho)c-gc^3~.
\label{eq:twoparam}
\end{eqnarray}
These equations employ the pitchfork bifurcation normal form with velocity
as the order parameter, coupled with the trivial observation that the velocity
is the rate of change of position.
The front velocity $c$ and position $x$, therefore constitute two degrees of 
freedom, that are sufficient to describe front dynamics close to the front
bifurcation.
The bifurcation parameter $\rho$ is denoted by $\gamma$ for the CGL equation
 and $\eta_c$ for the FN model. The two independent variables $c$ and $x$ in
Eq.~(\ref{eq:twoparam}), which are obviously uncoupled, 
represent dynamics where translational invariance is present, and the
solutions are independent of a choice of spatial origin. Scenarios can be
envisioned where this translational invariance is broken. Examples are the
imposition of different boundary conditions \cite{meron1}
or spatial
inhomogeneities introduced externally via an advective field 
\cite{meron4}.
To account for broken translational invariance in these scenarios,
 one has to modify Eq.~(\ref{eq:twoparam}).
 This would lead to the coupling
of the two degrees of freedom $x$ and $c$, the details of which 
would depend on the 
scenario considered. In this paper, we introduce a spatial
inhomogeneity in the form of Dirichlet boundary conditions,
examining its effect on incoming Bloch fronts  
in both models, and drawing parallels. The next section constitutes our
numerical study. In section IV, we derive and analyse the way 
in which Dirichlet boundary 
conditions couple $x$ and $c$ for the FN model. Such an analysis is not 
possible for the CGL equation since it is impossible to represent the
front position by a single point. Therefore, we rely exclusively on numerical 
simulations for that equation.

\section{Numerical Results}
In this section we lay out the numerical details of the study of the CGL and
FN models. The simulations in the two models are carried out in regimes where
analytical calculations performed by reducing front dynamics to a fewer
degrees of freedom are not possible. This is the regime where
Bloch fronts have high velocities (far beyond the front bifurcation 
threshold), and the fields forming the front core, have similar spatial scales.

We solve both the CGL and FN  system of reaction diffusion
equations using an implicit Crank--Nicholson scheme. Dirichlet
boundary conditions are used at both ends of the domain, which is typically
composed of $400$ grid points with a time step size of $0.01$. 
The boundary values
at one end are fixed at one of the homogeneous solutions of the
 Eq.~(\ref{eq:cgle}) and (\ref{eq:fitz}). 
At the other end we are free to vary the boundary condition.
In our numerical simulations,
we keep the domain large compared to the characteristic spatial extent
of the front, so that the influence of the boundary is only felt
when the front is close enough to it. We verified that the grid and
time steps were small enough to ensure that the numerical solution converged.

\begin{figure}
\begin{center}
\begin{tabular}{cc}
\resizebox{80mm}{!}{\includegraphics[width=6cm,height=8cm,angle=270]
{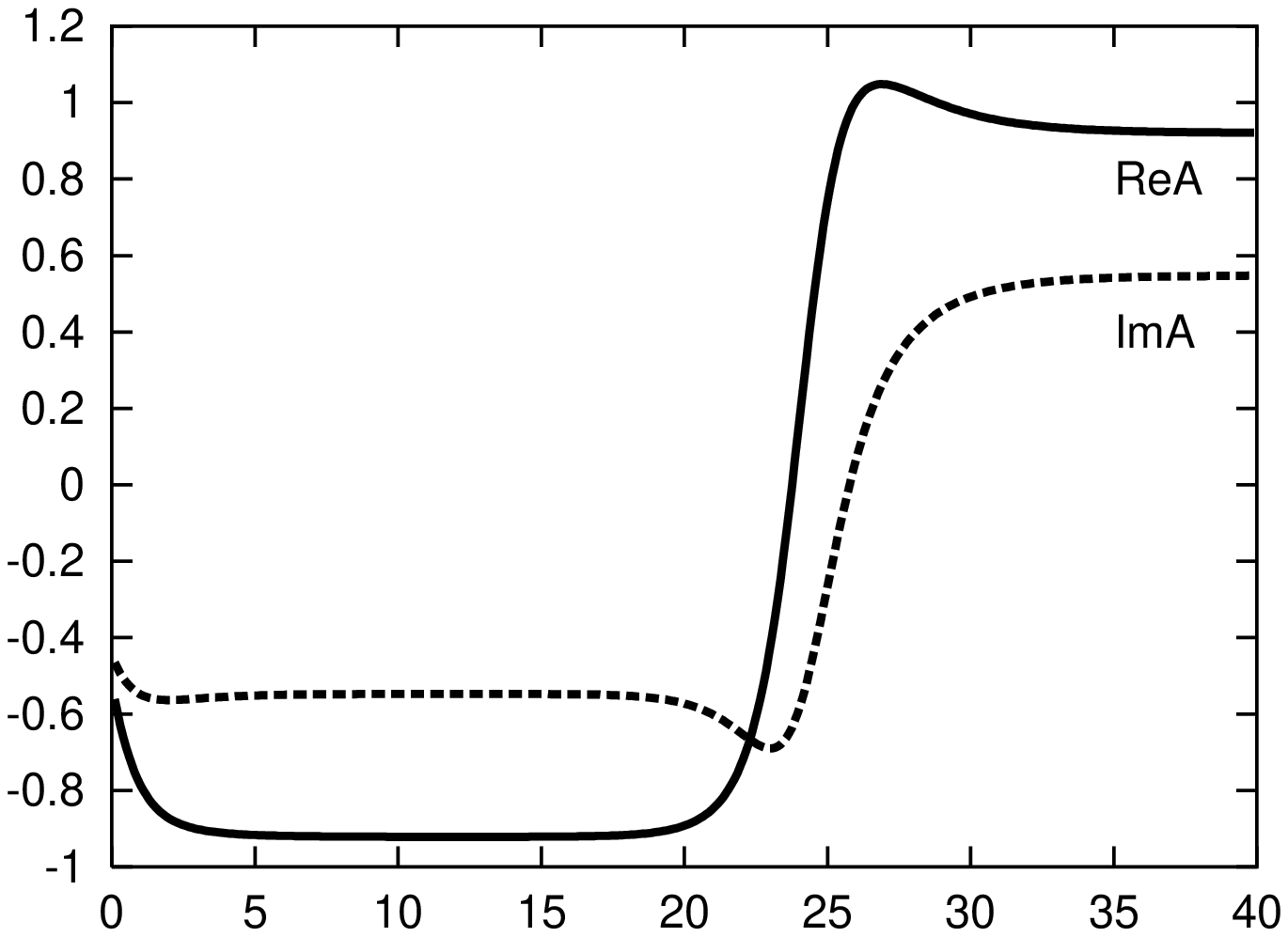}} \\
\resizebox{80mm}{!}{\includegraphics[width=6cm,height=8cm,angle=270]{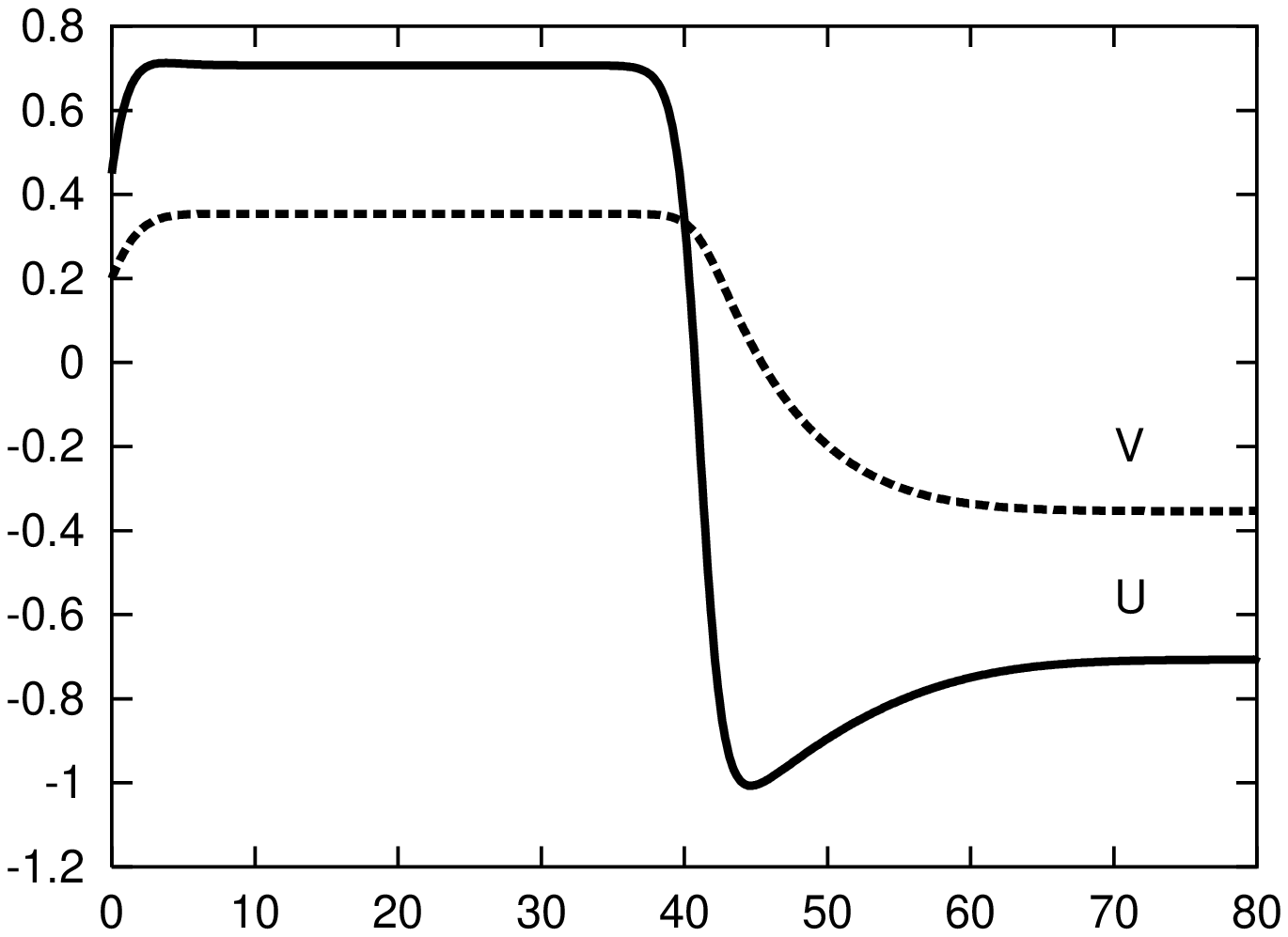}}
\end{tabular}
\caption{(a) Typical traveling front for the CGLE
(b) Traveling front in the FN model}
\label{fig:frontpic}
\end{center}
\end{figure}

By a suitable choice of
initial conditions a Bloch front or its counterpropagating partner
can be generated. The symmetry of the bifurcation ensures that they have
the same speed as long as they are not close to a boundary. A typical
front for the parametrically forced CGL equation and the FN system of
equations is shown in Fig.~\ref{fig:frontpic}. Bloch fronts for the CGL model
show a characteristic chirality broken
structure at their core \cite{Coullet} which is
essential for their propagation.
Similar structure considerations apply to fronts in the FN model
\cite{meron8}. 

\begin{figure}
\begin{center}
\begin{tabular}{cc}
\resizebox{80mm}{!}{\includegraphics[width=6cm,height=8cm,angle=270]
{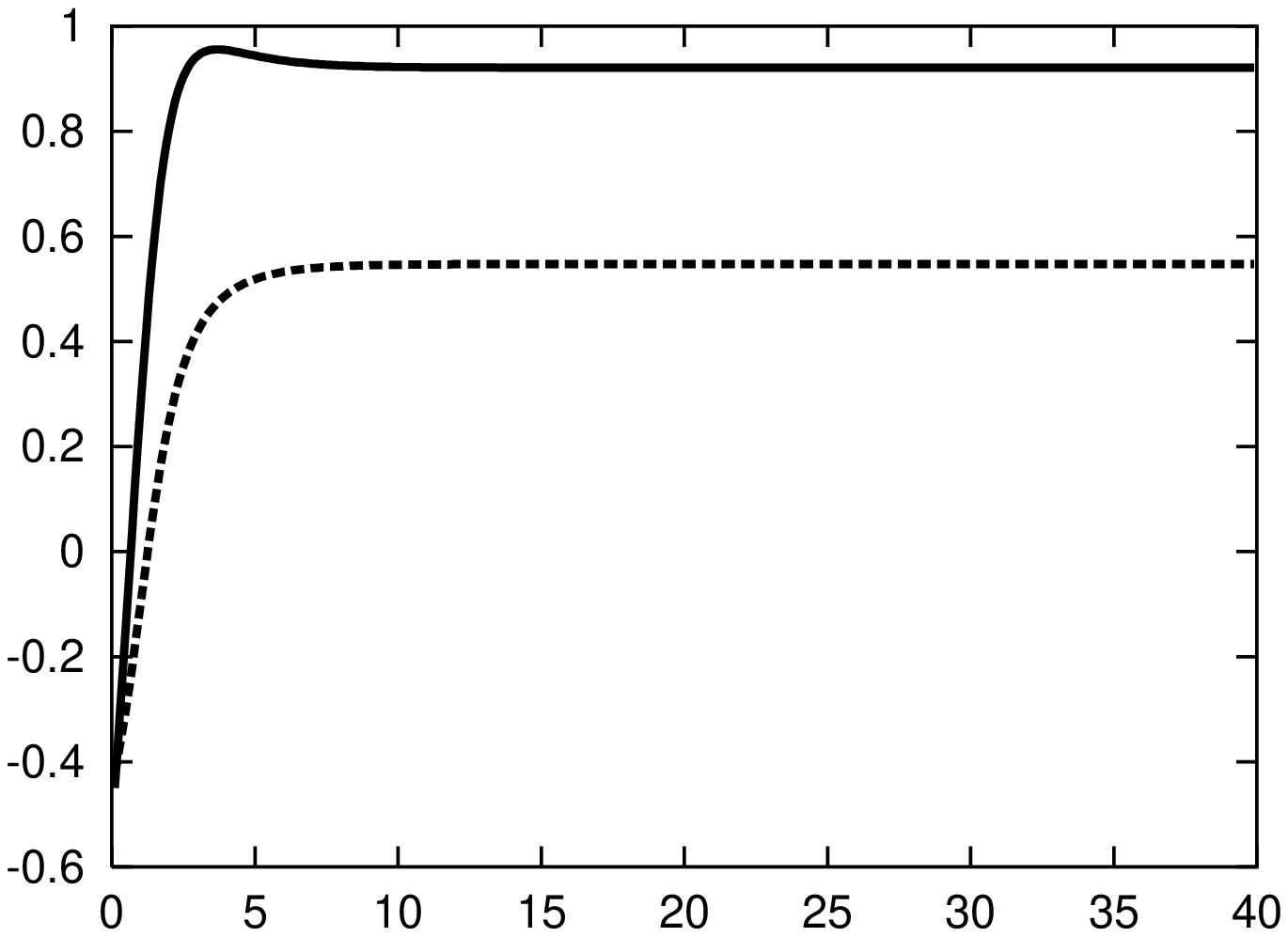}} \\
\resizebox{80mm}{!}{\includegraphics[width=6cm,height=8cm,angle=270]
{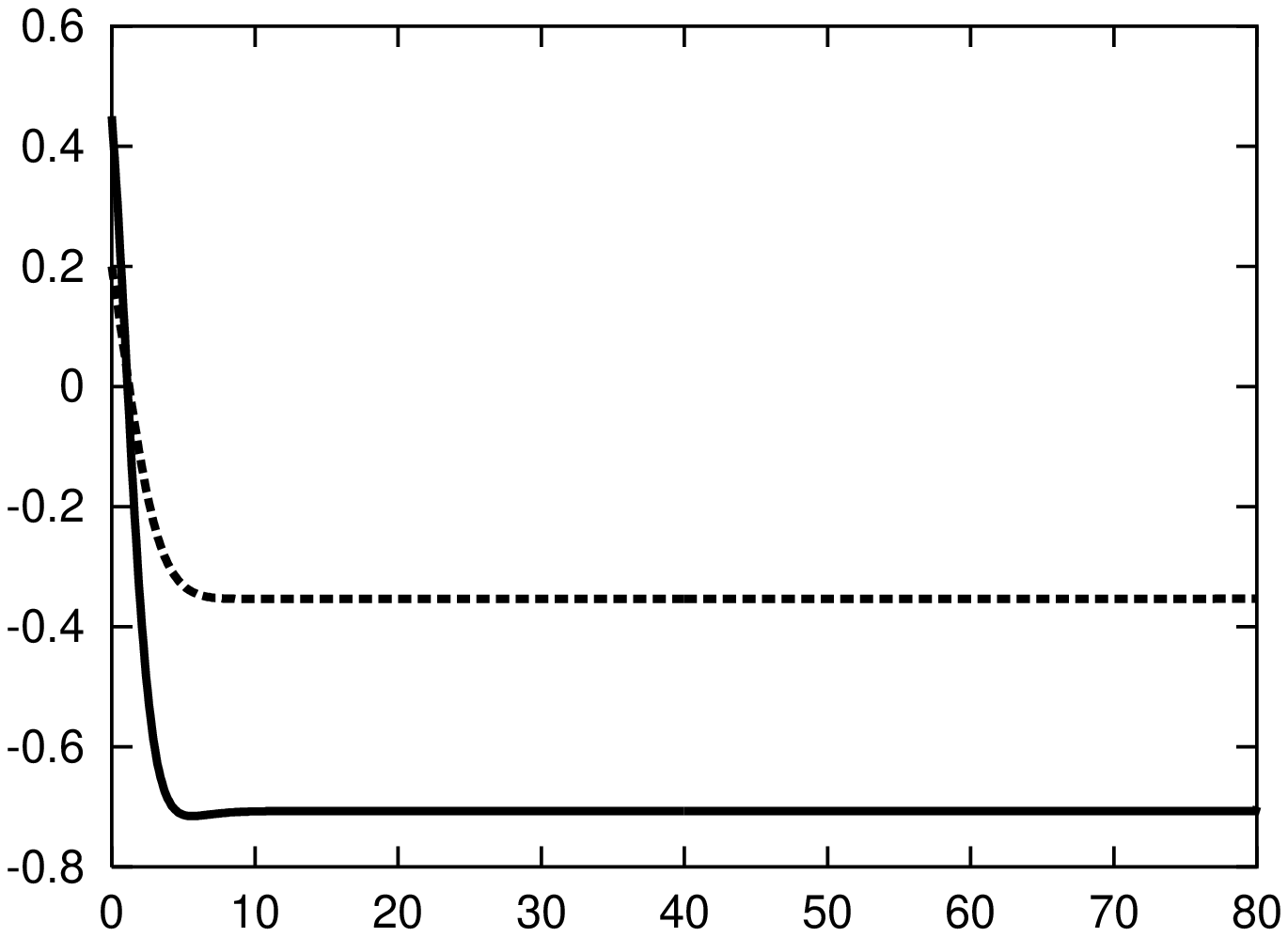}}
\end{tabular}
\caption{(a) A typical Nontrivial stationary solution to which
trapped fronts evolve into for the CGLE. The spatially
homogeneous solutions are connected to the Dirichlet boundary values.
(b) Nontrivial stationary solution for the FN model.}
\label{fig:nontr}
\end{center}
\end{figure}

In our simulations we focus on the interaction of incoming Bloch walls
 with boundaries, where Dirichlet boundary conditions are imposed on the
two fields, Re$A$, Im$A$ in the CGL equation and $u$, $v$ in the FN model.
Particularly we look at how the front cores are perturbed by the boundary
for a whole range of boundary conditions.
      Fronts coming into the boundary from infinity in both models 
either rebound 
or get trapped depending on the Dirichlet boundary values. A front 
that traps loses its core structure and evolves into 
the nearest available stable (attracting) configuration of the fields, 
which in this case is the nontrivial steady state solution of 
Eq.~(\ref{eq:cgle}) or Eq.~(\ref{eq:fitz}) for that particular boundary
condition. Figure~\ref{fig:nontr} shows a typical nontrivial steady state 
solution for the CGLE and FN model. This solution is comprised
 of a spatially homogeneous part, given by Eq.~(\ref{eq:trivcgle})
for the CGLE or Eq.
(\ref{eq:trivfitz}) for the FN model, and an inhomogeneous part 
that connects the
spatially homogeneous solutions to the Dirichlet boundary value. 
Rebounding phenomena close to the boundary is characterized by the core
of an incoming front flipping into the core of its counterpropagating partner,
resulting in the front moving away.

\begin{figure}
\begin{center}
\includegraphics[width=6cm,height=8cm,angle=270]{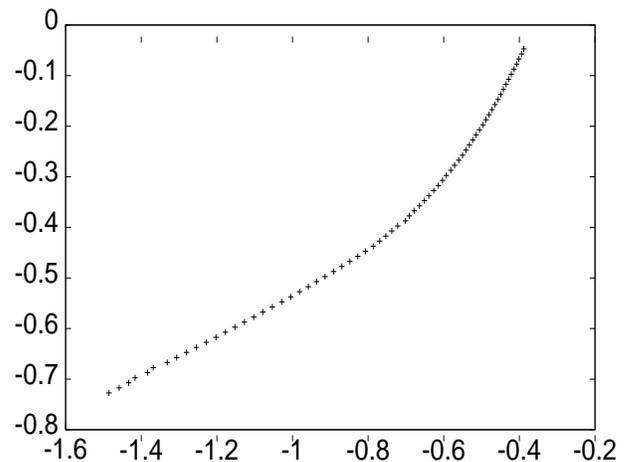}
\caption
{Transition Curve from a region of trapped Bloch fronts to bouncing ones
for the CGLE. ReA boundary value on the X axis, ImA boundary value
on the Y axis}
\label{fig:cgle}
\end{center}
\end{figure}

\begin{figure}
\begin{center}
\includegraphics[width=6cm,height=8cm,angle=270]{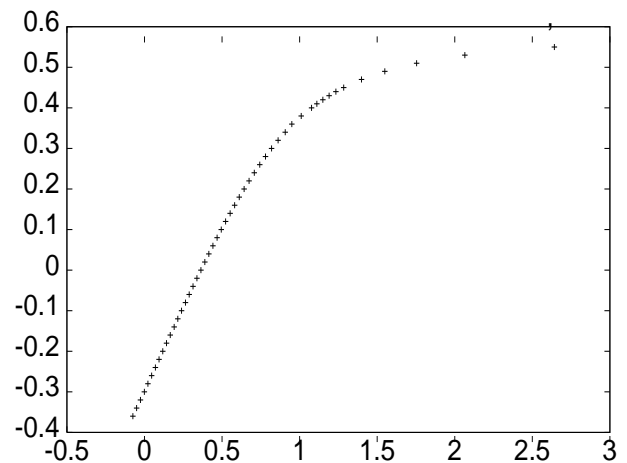}
\caption{\label{fig:fitz}
Transition Curve from a region of trapped Bloch fronts to bouncing ones
for the FN model. U boundary value on the X axis, V boundary value on the
Y axis.}
\end{center}
\end{figure}

Our observations are plotted in the plane of boundary values, revealing
a curve separating regions of bouncing and trapped fronts for both CGL and 
FN systems. 
Figure~\ref{fig:cgle} shows the
transition curve for $\alpha=-0.1$, $\gamma=0.31$, $\beta=-0.15$,
$\mu=1.0$ and $\nu=0.1$ for the parametrically forced CGL equation. A similar
transition curve for the FN model, with $a_1=2.0$,
$\delta=0.14$, $\epsilon=0.05$ is shown in Fig.~\ref{fig:fitz}. 
As one closes in on the curve 
from the trapping (bouncing) region, 
the fronts take longer to get trapped (bounce).
This slowing down in the dynamics close to the transition curve is indicative
of critical behavior, usually associated with eigenvalues of fixed points
that approach zero when a parameter (the Dirichlet boundary value here)
 is varied. An analytical explanation
of the slowing down is given in the next section.

Incoming Bloch fronts, in both the CGL and FN models, evolve into nontrivial
stationary solutions to Eq.~(\ref{eq:cgle}) and Eq.~(\ref{eq:fitz})
respectively when trapped at the boundary. These nontrivial solutions
are linearly stable by virtue of the Bloch fronts evolving into them. It
remains to be ascertained whether these solutions remain stable when
Dirichlet boundary values that lead to a bounce are imposed at the boundary.
Hypothetically, one could associate the loss of stability of these nontrivial
solutions with the transition from trapping to bouncing of incoming fronts. 
We test this hypothesis by carrying out a numerical linear stability 
analysis of the nontrivial stationary solutions $\psi_o$.
Eq.~(\ref{eq:cgle}) is linearized to,
\begin{eqnarray}
{\partial \delta\psi\over \partial t} & = &
(1+i\alpha)\nabla^2\delta \psi +
\lbrack\mu+i\nu - 2(1+i\beta)|\psi_o|^2\rbrack\delta\psi\nonumber\\
& &
+ \lbrack\gamma - (1+i\beta)\psi_o^2\rbrack\delta\psi^{*}~,
\label{eq:lin}
\end{eqnarray}
for the perturbation $\delta\psi=0$ at the boundaries. A similar
linearization is done for the FN system. It is found that
the eigenvalue spectrum for both systems has negative real parts in the
trapping region, as expected.
Moreover, these real parts remain negative when we evaluate the stability
of the nontrivial stationary solutions in the bouncing region. Even 
though nontrivial 
solutions are stable in the bouncing region, incoming fronts
do not evolve into them. Consequently, the critical behavior is not 
governed by the loss of stability of these solutions.

We also observe that close to the transition curve bouncing and
trapped fronts can coexist.
Instead of coming in from infinity and rebounding,
a Bloch front created close to the boundary may get trapped even if 
Dirichlet boundary values that produce a bounce are imposed.
To demonstrate this,
we choose a boundary value inside the trapping 
region, close to the transition curve.
A Bloch wall is launched from infinity towards the boundary. As the wall
approaches the boundary, the simulation is stopped. The field
configuration is saved and used as the initial condition for the next
simulation run. In this new run we make a small change in the Dirichlet
boundary value and move across the transition curve into the bouncing region.
If the front core in the saved configuration is 
close enough to the boundary, the front will get trapped, even if
Dirichlet boundary values that produce a bounce are imposed.
An analytical explanation of coexistence
for the FN model is given in the next section, which gives
insights into the coexistence behavior in both models.

Summarizing, in both systems we have regions of trapped incoming Bloch
fronts and bouncing Bloch fronts in the plane of boundary values.
Critical slowing down of the front dynamics in proximity to a 
boundary is observed in both systems close to the transition curve.
Also, the nontrivial stationary solutions in both systems remain stable
across the transition curve, implying that they are not 
responsible for the critical behavior we see. Both systems exhibit the
coexistence of bouncing and trapped solutions.

In the next section, we explain these numerical observations
by deriving the mechanism that shows how $x$ and $c$ are coupled for
Dirichlet boundary conditions in the FN model.

\section{Analysis}

The goal of this section is to explain front interaction with boundaries,
 in terms of coupling of the
evolution equations for the two degrees of freedom, front velocity $c$ 
and the front position $x$ in Eq.~(\ref{eq:twoparam}). It
is shown that the coupling is the result of the spatial inhomogeneity
sensed by the front as it approaches the boundary.

 We restrict ourselves to the regime where $\epsilon/\delta$ is small
and $\eta=\sqrt{\epsilon\delta}$ is finite. This restriction leads to
a very sharp spatial variation of $u(x,t)$ field at the core of the front. The
slowly varying $v(x,t)$ field can be considered constant in this
core region. Hence the $v(x,t)$ field is represented by
a single value $v_f$, at the point where
the $u(x,t)$ field has zero value. We also restrict ourselves to small
front velocities $c$, which can be done, by either making $\eta_c-\eta$
small or choosing a large $a_1$.

The restrictions described above lead to a 
drastic reduction in the
number of degrees of freedom used to describe the front. As opposed to
a front description based on the whole model Eq.~(\ref{eq:fitz}),
 a sharp front (fast spatial variation of $u$ at the core)
can be thought of as a point particle with a definite position and velocity.
The slowly varying $v$ variable can be thought of as a field associated
with this particle that allows it to sense the boundary. A small velocity
has a few simple implications. An addition of two kinds of perturbations,
one that changes $v_f$ (velocity) slightly and the other 
that produces a local distortion in the $v(x,t)$ field far from the 
front position, to a slow moving and uniformly translating front, is 
followed by the disappearance of the distortion and the relaxation of 
the front back to constant velocity. The time scale on which the distortion
vanishes is much faster than the scale on which $v_f$ relaxes back to 
its original value. 
In essence, we have two time scales, the slower one associated with
nonsteady front motion.

We now employ the restrictions mentioned above
to obtain a reduced  description of Eq.~(\ref{eq:fitz}). 
We solve Eq.~(\ref{eq:fitz}) with
a Dirichlet boundary condition $v=v_b$ at the left boundary. On the 
right boundary, which is at infinity, we have $v=-q^{-2}$, and
$q^2=a_1+1/2$.
Following Ref.~\cite{meron4}, we have the following system of equations,
\begin{eqnarray}
\dot{x}=-\frac{3}{\eta \sqrt{2}}v_f,
\label{eq:velocity}
\end{eqnarray}
and
\begin{eqnarray}
v_t+q^{2}v-v_{rr}&=&-\frac{3}{\eta \sqrt{2}}v(0,t)v_r+1~~~~~r\leq 0\nonumber\\
v_t+q^{2}v-v_{rr}&=&-\frac{3}{\eta \sqrt{2}}v(0,t)v_r-1~~~~~r\geq 0\nonumber\\
v(-x,t)=v_b&,\quad&v(\infty,t)=-q^{-2}.
\label{eq:system}
\end{eqnarray}
Eq.~(\ref{eq:velocity}) implies that the velocity is proportional to
the value of the $v(x,t)$ field at the sharp interface formed by the $u(x,t)$
field. Eq.~(\ref{eq:system}) represents the evolution of the $v(x,t)$ field
in a frame of reference which moves with the front. The variable
 $r$ is the spatial coordinate from the front position
 and $r=-x$ is the distance of the boundary to the left of the front.

We solve Eq.~(\ref{eq:system}) perturbatively. The starting point
of the perturbative expansion is to find a stationary Ising wall solution.
Therefore, setting the time derivatives in Eq.~(\ref{eq:system}) to zero
 and looking for an Ising wall solution, we get,
\begin{eqnarray}
v_{rr}-q^{2}v +1&=&0 ~~~~~r\leq 0,\nonumber\\
v_{rr}-q^{2}v -1&=&0 ~~~~~r\geq 0,
\label{eq:isingsol}
\end{eqnarray}
with $v(0_+)=v(0_-)=0$ and $v(\infty)=-q^{-2}$.  This ensures that
that the Ising wall has $v_f=0$. The solution to 
Eq.~(\ref{eq:isingsol}) is,
\begin{eqnarray*}
v^{(0)}&=&-q^{-2}(e^{qr}-1)~~~r\leq 0,\nonumber\\
v^{(0)}&=&q^{-2}(e^{-qr}-1)~~~r\geq 0.
\end{eqnarray*}
Hence for the Ising wall at $r=-x$, we have,
$v^{(0)}(-x)=(1-e^{-qx})/q^{2}$. 

We look for traveling Bloch wall solutions as a perturbation to this
uniquely defined Ising wall. Since the Bloch walls have a Dirichlet 
boundary condition $v(-x)=v_b$, the perturbative correction to the Ising wall
should have a boundary value $v_c=v_b-(1-e^{-qx})/q^{2}$ which changes as 
the front moves.

Let $\overline{v}$ be the perturbation. Then,
\begin{equation} 
v=\overline{v}+v^{(0)}.
\label{eq:lala}
\end{equation}
$\overline{v}$ is expanded in powers of $c$, the small perturbation parameter.
Since the front bifurcation is a pitchfork, $\eta$ is expanded in powers of
$c^{2}$.
\begin{eqnarray}
\overline{v}(r,t,T)&=&\sum_{n=1}^{\infty}c^{(n)}v^{(n)}(r,t,T)\nonumber\\
\eta&=&\eta_c(x)-c^{2}\eta_1(x)+c^{4}\eta_2(x).
\label{eq:pert}
\end{eqnarray}
$T=c^{2}t$ is the slower time scale responsible for nonsteady front motion.
The coefficients of powers of $c$ in the expansion of $\eta$ are functions
of $x$ to incorporate the broken translational invariance.
Using Eq.~(\ref{eq:lala}) and Eq.~(\ref{eq:pert}) in Eq.~(\ref{eq:system}),
one obtains
\begin{equation}
v_{t}^{(n)}+q^{2}v^{(n)}-v_{rr}^{(n)}=-\rho^{(n)},~~~~n=1,2,3..,
\label{eq:lindif}
\end{equation}
with
\begin{eqnarray}
\rho^{(1)}&=&\frac{3}{\sqrt{2}\eta_c}v_{|r=0}^{(1)}v_{r}^{(0)}\nonumber\\
\rho^{(2)}&=&\frac{3}{\sqrt{2}\eta_c}
\left[v_{|r=0}^{(1)}v_{r}^{(1)} +
v_{|r=0}^{(2)}v_{r}^{(0)}\right]\nonumber\\
\rho^{(3)}&=&v_{T}^{(1)}+\frac{3\eta_1}{\sqrt{2}\eta_c^2}
v_{|r=0}^{(1)}v_{r}^{(0)}\nonumber\\
&+&\frac{3}{\sqrt{2}\eta_c}\left[
v_{|r=0}^{(1)}v_{r}^{(2)}+v_{|r=0}^{(2)}v_{r}^{(2)}+
v_{|r=0}^{(3)}v_{r}^{(0)}\right].
\label{eq:pertex}
\end{eqnarray}
We use Green's functions to solve the system of equations above. The general
solution, given that we have found an appropriate Green's function 
$G(r,t|r^{\prime},t^{\prime})$ is,
\begin{eqnarray}
v^{(n)}(r,t)&=&
\int_{t_i}^{t}\int G(r,t|r^{\prime},t^{\prime})\rho^{n}(r^{\prime},t^{\prime})\,
dt^{\prime}\,dr^{\prime}\nonumber\\
&+&\int_{t_i}^{t}\int_{s^{\prime}}\left[G(r,t|r^{\prime},t^{\prime})
\frac{\partial^{\prime}{v^{(n)}(r^{\prime},t^{\prime})}}{\partial n}\right.
\nonumber\\
&-&\left.\frac{\partial^{\prime}{G(r,t|r^{\prime},t^{\prime})}}
{\partial n}v^{(n)}(r^{\prime},t^{\prime})\right]
\,dt^{\prime}\,ds^{\prime}\nonumber\\ 
&+&\int G(r,t|r^{\prime},t_{i})v^{(n)}(r^{\prime},t_i)\,dr^{\prime}.
\label{eq:green}
\end{eqnarray}
The last term can be made zero by choosing an appropriate initial condition.
The first term gives the influence of sources on the evolution of the $v(x,t)$
field. The second term incorporates the influence of boundary conditions.
To apply Dirichlet boundary conditions one finds a Green's function which
is zero at the left boundary.
Since we have a semi-infinite domain, we use the method of images
to write down the Green's function
\begin{eqnarray}
& & G(r,t|r^{\prime},t^{\prime})=\frac{e^{-q^2(t-t^{\prime})}}
{\sqrt{4\pi(t-t^{\prime})}}
\exp\left[-\frac{{(r-r^{\prime})}^2}{4(t-t^{\prime})}\right]\nonumber\\
& & -\frac{e^{-q^2(t-t^{\prime})}}{\sqrt{4\pi(t-t^{\prime})}}
\exp\left[-\frac{{(r+r^{\prime}+2x(t^{\prime}))}^2}{4(t-t^{\prime})}\right],
\label{eq:funct}
\end{eqnarray}
where the second term is the image of the first and $G=0$ at $r=-x$,
the boundary.
Proceeding with the calculation of source effects in Eq.~(\ref{eq:green}),
we have,
\begin{eqnarray}
&&v^{(n)}(r,t)= 
 \int_{-x(t^{\prime})}^{\infty}\int_{0}^{t}
\frac{e^{-q^2(t-t^{\prime})}}{\sqrt{4\pi(t-t^{\prime})}}\left\{
\exp\left[-\frac{{(r-r^{\prime})}^2}{4(t-t^{\prime})}\right]\right.
\nonumber\\&&-\left.
\exp\left[-\frac{{(r+r^{\prime}+2x(t^{\prime}))}^2}{4(t-t^{\prime})}\right]
\right\}\rho^{(n)}
(r^{\prime},t^{\prime})\,dt^{\prime}\,dr^{\prime}.
\label{eq:source}
\end{eqnarray}
The Green's function terms in Eq.~(\ref{eq:source}) above contain exponentials
of functions of time $t^{\prime}$ that possess a maxima at some time 
$t^{\prime}_o$. 
Hence most of the contribution to the integral comes around this maximum
 value, which is
approximately given by $t-t^{\prime}_o=|r-r^{\prime}|/2q$ for the first integral,
and $t-t^{\prime}_o=|r+r^{\prime}+2x(t^{\prime}_o)|/2q$ for the second. If the
 width of the maxima peak is less than these time differences, we can take the 
limit $t\rightarrow\infty$ in the integrals above. Further, if all the source 
terms are smoothly varying functions of $t^{\prime}$, one could perform a
steepest descent approximation to the integrals above assuming that
the maxima peak is nearly a gaussian and sharp enough. 
Physically, as a consequence of both reaction and diffusion,
the configuration of fields at present time $t$ in Eq.~(\ref{eq:lindif})
is determined by the time behavior of sources in an earlier small time 
window, in which reaction and diffusion mechanisms combine to produce
the maximum rate of change of the $v(x,t)$ field. At all other times, either
reaction or diffusion is individually dominant and not able to produce 
a combined high rate of change of $v(x,t)$.
This is unlike pure diffusion, where all the time history of sources 
is required to give the field configuration at present time.
Performing the steepest descent calculation,
the time portion of the integral in Eq.~(\ref{eq:source}) can be eliminated
and it reduces to,
\begin{equation}
v^{(n)}(r)=\int_{-x}^{\infty}\left[g+ \frac{pe^{-(r+r^{\prime}+2x)f}}
{\sqrt{2}\sqrt{a(r+r^{\prime}+2x)+b}}\right]\rho^{(n)}(r^{\prime})\,dr^{\prime},
\label{eq:resource}
\end{equation}
where,
\begin{eqnarray*}
       g & = & {e^{-q|r-r^{\prime}|}}/{2q},\\
       p & = & ({\dot{x}\pm\sqrt{\dot{x}^2+q^2}})/{2q^2},\\
       f & = & q^2 p +1/4p,\\
       a & = & \ddot{x}p^2,\\
       b & = & 2\dot{x}^2 p^2 +2 \dot{x} p +1/2.
\end{eqnarray*}
The first factor in the bracelets $g$ dominates the evolution far away
from the boundary, while the second factor is responsible for sensing
the boundary.
 One notices that $\dot{x}$ and $\ddot{x}$, the velocity and acceleration
of the front are involved in the reduced Green's functions in 
Eq.~(\ref{eq:resource}). 
If $\dot{x}$ and $\ddot{x}$ are neglected in the expressions above (justifiably
so since we are close to the front bifurcation), solving Eq.~(\ref{eq:lindif})
reduces to solving 
\begin{equation}
q^{2}v^{(n)}-v_{rr}^{(n)}+\rho^{(n)}=0,~~~~n=1,2,3..
\end{equation} This is what the authors do in Ref.~\cite{meron4}, although no
boundary influence is considered.
Solving Eq.~(\ref{eq:resource}) further for $n=1,2,3$, and requiring the
smoothness of the front at all orders, we get the equation for the evolution of
$v_f=cv^{(1)}$.
\begin{eqnarray}
\dot{v_f}&=&-\frac{v_f}{\sigma}\left[\frac{\partial\sigma}{\partial x}
\frac{\partial x }{\partial t} + \frac{\partial\sigma}{\partial \dot{x}}
\frac{\partial\dot{x}}{\partial t} +\frac{\partial\sigma}{\partial \ddot{x}}
\frac{\partial\ddot{x}}{\partial t}\right]\nonumber\\
&+&\frac{4q^2\sigma\sqrt{2}}{9}(\eta_c-\eta)
v_f-\frac{\sigma^2}{6}v_f^3,
\label{eq:mero}
\end{eqnarray} where,
\begin{eqnarray*}
\sigma &=& 3/\sqrt{2}\eta_c\\
       &=& \left[(2-e^{-2qx})/4q^{3}
+S_1+S_2\right]^{-1},
\end{eqnarray*}
 and
\begin{eqnarray*}
S_1=\int_{-x}^{0}\frac{pe^{-(r+r^{\prime}+2x)f}}{\sqrt{2}
\sqrt{a(r+r^{\prime}+2x)+b}}
\left[\frac{-e^{qr^{\prime}}}{q}\right]\,dr^{\prime},
\end{eqnarray*}
\begin{eqnarray*}
S_2=\int_{0}^{\infty}\frac{pe^{-(r+r^{\prime}+2x)f}}
{\sqrt{2}\sqrt{a(r+r^{\prime}+2x)+b}}\left[\frac{-e^{-qr^{\prime}}}
{q}\right]\,dr^{\prime}.
\end{eqnarray*}
Hence $S_1$, $S_2$ and so $\sigma$ depend on $x$, $\dot{x}$, $\ddot{x}$. 
Neglecting $\dot{x}$ and $\ddot{x}$ in $S_1$ and $S_2$, $\sigma$ reduces to,
\begin{equation}
\sigma=\frac{4q^{3}}{2+(1+2qx)e^{-2qx}}.
\end{equation}
 Far away from the boundary influence, one recovers $\sigma=2q^3$
as expected. Therefore, one of the effects of proximity to the boundary is 
spatial dependence of the critical bifurcation parameter $\eta_c$. Due to 
this form of $\sigma$, where its dependence on velocity and acceleration 
is ignored and its spatial derivative falls off sharply with $x$,
all its derivatives in Eq.~(\ref{eq:mero}) can be neglected. Thus we have,
\begin{equation}
\dot{v_f}=\frac{4q^2\sigma\sqrt{2}}{9}(\eta_c-\eta)
v_f-\frac{\sigma^2}{6}v_f^3.
\label{eq:mero1}
\end{equation}

We now derive the effect of boundary conditions on the evolution of the
front. Taking the derivative of $G$ with respect to $r^{\prime}$ in 
Eq.~(\ref{eq:funct}) and
substituting it in Eq.~(\ref{eq:green}), the boundary contribution
is found to be
\begin{eqnarray}
\phi_1 &=& -\int_{0}^{t}\frac{v_{c}\textbf{[}x(t^{\prime})\textbf{]}
e^{-q^2(t-t^{\prime})}}{{2\pi}^{1/2}
{(t-t^{\prime})}^{3/2}}\nonumber\\ &\times&
[r+x(t^{\prime})]\exp\left[-\frac{{[r+x(t^{\prime})]}^2}{4(t-t^{\prime})}\right]
\,dt^{\prime}.
\end{eqnarray}
Now we are only interested in the contribution of the boundary terms at 
the front position, $\phi_1|_{r=0}$. This extra term gets added on to 
$v_f$, the value of the $v$
 field at the front position, thus incorporating the influence
of the specific Dirichlet boundary condition $v_b$ on the front velocity.
From now on $\phi_1$ will stand for $\phi_1|_{r=0}$.
 
Nonsteady front motion represented by $\dot{v_f}$ involves time derivatives,
 hence the time derivative of $\phi_1$ gives the influence of the boundary
condition in accelerating fronts,
\begin{eqnarray}
\phi_2&=&\frac{\partial{\phi_1}}{\partial{t}}\nonumber\\
&=&-\int_{0}^{t}
\frac{v_{c}\textbf{[}x(t^{\prime})\textbf{]}e^{-q^2(t-t^{\prime})}}
{{2\pi}^{1/2}{(t-t^{\prime})}^{3/2}}
\exp\left[-\frac{{[x(t^{\prime})]}^2}{4(t-t^{\prime})}\right]
\nonumber\\&\times &\left(
q^2x(t^{\prime})+\frac{3x(t^{\prime})}{(t-t^{\prime})}
-\frac{{[x(t^{\prime})]}^3}{{4(t-t^{\prime})}^2}
\right)\,dt^{\prime}.
\end{eqnarray}

Incorporating these boundary effects in Eq.~(\ref{eq:mero1}),
$v_f       \rightarrow  v_f+\phi_1$, and
$\dot{v_f} \rightarrow \dot{v_f}+\phi_2$, we get
\begin{eqnarray}
\dot{v_f}=\phi_2+\frac{4q^2\sigma\sqrt{2}}{9}(\eta_c-\eta)
(v_f+\phi_1)-\frac{\sigma^2}{6}{(v_f+\phi_1)}^3.\nonumber\\
\quad
\label{eq:mero2}
\end{eqnarray} 
Eq.~(\ref{eq:velocity}) and Eq.~(\ref{eq:mero2}) 
constitute the coupling of the two degrees of freedom
$c$ and $x$ or equivalently $v_f$ and $x$ in the presence of a spatial
inhomogeneity introduced by Dirichlet boundary conditions. 

One of the stationary points of the system above is 
$(x,v_f)=\left({-\ln(1-v_bq^2)}/{q},0\right)$. This is the point
closely connected with the dynamics we now describe. We look at the
parameter $v_b$ in three different regimes, $v_b<0$, $0 <v_b<1/q^2$,
and $v_b>1/q^2$. The region of physical interest is $x>0$. $v_f>0$ 
implies a front moving towards the boundary and $v_f<0$ is a front moving away.

Although, the convolution integrals $\phi_1$ and $\phi_2$ represent the 
complete influence of boundary conditions, we wish to approximate them
to obtain a simpler picture that preserves the 
qualitative features of the complete integrals. The integrals involve
exponentials of functions that have a maxima at time $t^{\prime}_o$,
and hence most of the contribution is around this maxima. For small front
velocities this maxima is given by $t-t^{\prime}_o=x/2q$. The width of this
maxima peak is given by $(x/2q)^{\frac{3}{2}}$. If $x/2q>>(x/2q)^{\frac{3}{2}}$,
which it is for a very sharp peak, a steepest descent approximation
can be made. A greater inequality implies a better approximation. 
The approximation gives,
$ \phi_1=-v_c(x)e^{-qx}$ and
$ \phi_2={\partial{\phi_1}}/{\partial{t}}=-3\phi_1q/x$,
where it is again assumed that the velocities are small.

\begin{figure}
\includegraphics[width=6cm,height=8cm,angle=270]{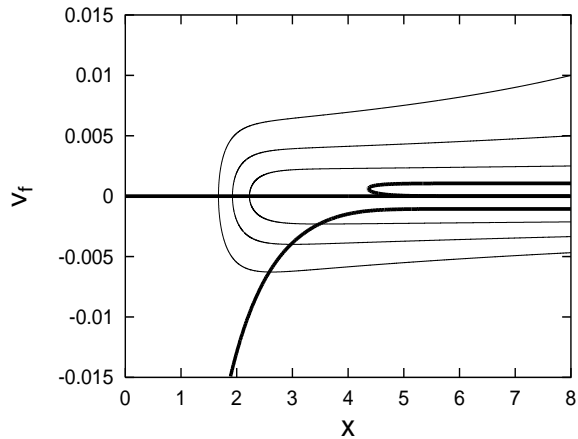}
\caption{The dark curves are the nullclines, grey curves are the
solutions to Eq.~(\ref{eq:mero2}) with different initial conditions.
These are the typical flows in the regime $v_b<0$, where incoming fronts
always bounce. There are no fixed points and $v_b=-0.1$,
$a1=9.9$, $\epsilon=.001$, $\delta=1.0$.}
\label{fig:bouncing}
\end{figure}

In the regime $v_b<0$ the fixed point is in the negative $x$ region and is
an unstable spiral. The term involving $1/x$ in the expression for $\phi_2$
prevents flows from crossing the negative $x$ region to the positive one and
vice versa. Therefore the fixed point does not influence the $x>0$ flows.
Figure~\ref{fig:bouncing} shows the nullclines and the typical flows. The grey
flow curves show the turning around of fronts at the boundary. It is notable
that the nullclines are not followed well by the flow curves and cannot
predict the dynamics of front reversal. Generally, the flows will agree better
with the nullclines when the relaxation rate, determined by
$\eta_c-\eta$, is larger, although the jump from one nullcline to another can 
only be explained by the dynamical equations.

As one increases $v_b$ and enters the $0<v_b<1/q^2$ regime, the fixed
point crosses
over into the $x>0$ region. The fixed point now influences the flows 
close to the boundary. Instead of being an unstable spiral it now is a saddle
with two distinct real
eigenvalues, giving rise to stable and unstable manifolds.
The eigenvalues are
\begin{equation}
\lambda_{\pm}=\frac{\lambda_o \pm \sqrt{\lambda_o^2 +4\lambda_x}}{2},
\label{eq:eigen}
\end{equation} where
$\lambda_o={4q^2\sigma\sqrt{2}}/{9}(\eta_c-\eta)$, and
\begin{equation}
\lambda_x=-\frac{(1-v_bq^2)}{\sqrt{2}\eta}\left[\frac{9q}
{\ln{(1-v_bq^2)}}\right].
\end{equation}
As $v_b\rightarrow1/q^2$, the fixed point moves away from the boundary
towards positive infinity and $\lambda_{+} \rightarrow \lambda_o$, which is 
the eigenvalue for an unstable Ising wall far from the boundary influence.
Also, in the same limit,
 $\lambda_{-}\rightarrow0$, where the zero eigenvalue is associated with 
spatial homogeneity (translational invariance). This explains the critical
slowing down observed in the last section. Trajectories wandering close to 
this fixed point near criticality ($v_b\rightarrow1 /q^2$) will rebound 
or trap on a slower time scale, compared to a relatively faster dynamics
when the fixed point is further away from criticality.

The time scales close
to the fixed point are controlled by $\lambda$, which has two constituents,
 $\lambda_o$ and $\lambda_x$. $\lambda_o$ is associated with the slow
time scale $T=c^2t$, which depends on the distance to the front 
bifurcation $\eta_c-\eta$, and can be made arbitrarily small. 
Therefore, close to the
front bifurcation, $\lambda_o$ can be neglected in Eq.~(\ref{eq:eigen}) and
the eigenvalues reduce to $|\lambda_{\pm}|=\sqrt{\lambda_{x}}$. This is the
new time scale determined solely by the influence of boundary conditions
and is the dominant time scale in the nonadiabatic limit of extremely slow
velocities.
Bloch wall trajectories close to the saddle, which either trap or bounce, 
evolve on this time scale.

\begin{figure}
\includegraphics[width=6cm,height=8cm,angle=270]{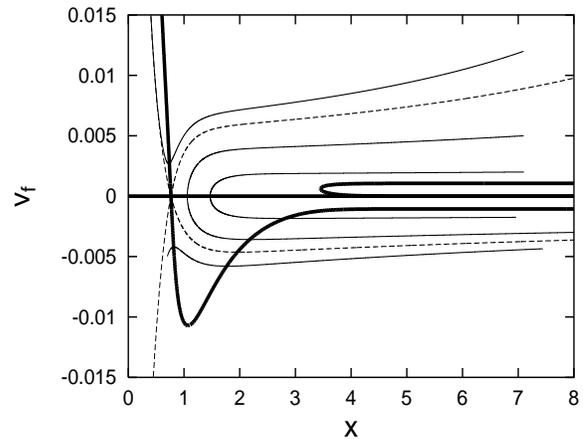}
\caption{Flows in the coexistence region $0<v_b<1/q^2$. Dark curves
are the nullclines, which intersect at the fixed point. Grey curves
are the solutions to Eq.~(\ref{eq:mero2}), with different initial conditions.
The dashed lines represent the invariant manifolds separating basins of
attraction and bouncing. $v_b=0.088$,
$a1=9.9$, $\epsilon=.001$, $\delta=1.0$.}
\label{fig:crit}
\end{figure}

\begin{figure}
\includegraphics[width=6cm,height=8cm,angle=270]{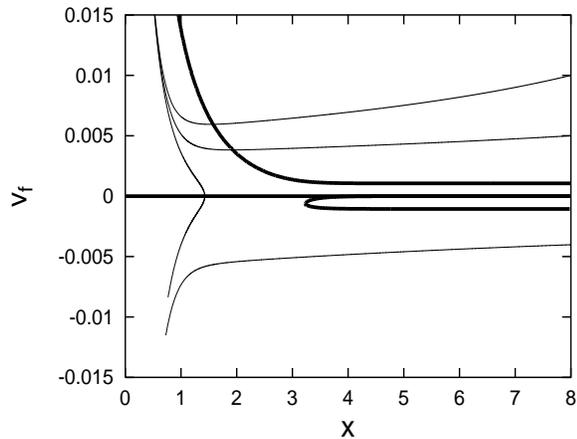}
\caption{Flows in the regime $v_b>1/q^2$. Again there are no
fixed points present and all the flows get trapped at the boundary.
 $v_b=0.1$, $a1=9.9$, $\epsilon=.001$, $\delta=1.0$}
\label{fig:trap}
\end{figure}

Typical flows are plotted 
in Fig.~\ref{fig:crit}. The triangle shows the fixed point. The dashed lines
are the invariant sets, with arrows showing the direction of flow. The 
invariant sets separate basins of attraction of flows towards the boundary
and basins of reflection away from it. This explains the coexistence region.
If the initial velocity and position of the front is inside the attraction
basin, it gets trapped at the boundary, if not, it rebounds.
The numerically obtained transition curve in the last section is a 
manifestation of this behavior, where initial conditions are fixed and 
varying the boundary values leads to a crossover of the initial condition
from a basin of attraction to that of repulsion. It should be noted that
the coexistence region can only be explained by the presence of the fixed
point and the dynamics associated with it. An adiabatic analysis relying
on nullclines is not the complete picture.

\begin{figure}
\includegraphics[width=6cm,height=8cm,angle=270]{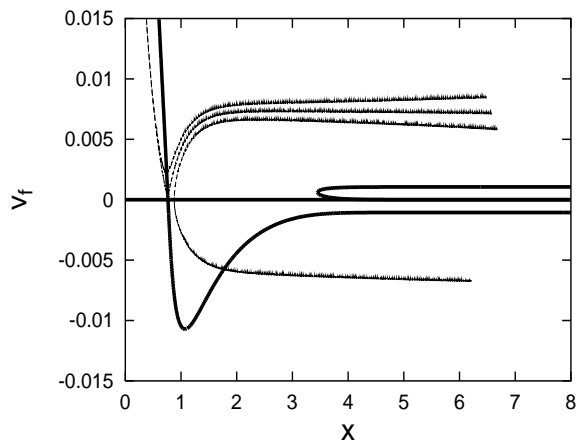}
\caption{This shows the actual simulation of Eq.~(\ref{eq:fitz}) for
$v_b=0.088$, $a1=9.9$, $\epsilon=.001$, $\delta=1.0$ 
The thick dark lines are the nullclines. The thin lines are the
trajectories.
}
\label{fig:actual}
\end{figure}

For $v_b>1/q^2$ the fixed point no longer exists. Figure~\ref{fig:trap} shows
the flows in this regime. Incoming fronts always get trapped.
Since no fixed points are present, the flow qualitatively does what the 
nullclines do, as is the case in the $v_b<0$ regime.

Summarizing, transition from bouncing to trapped fronts is governed by
a fixed point close to the boundary. This fixed point gives rise to the
coexistence behavior, and is absent in regimes where only trapping or
bouncing occurs. We conclude our analysis by pointing out that solutions of
Eq.~(\ref{eq:mero1}), with approximated $\phi_1$ and $\phi_2$,
agree well qualitatively with the solution of the FN model 
Eq.~(\ref{eq:fitz}) plotted in Fig.~\ref{fig:actual}.

\section{Conclusion}

We have studied Bloch front motion in the CGL and FN models
in the presence of spatial inhomogeneity introduced by
Dirichlet boundary conditions at the boundary. We have
shown similar features in the behavior of Bloch fronts close
to the boundary in both systems.
There is a transition from trapping (annihilation) to bouncing
of incoming Bloch fronts for both systems as a function of Dirichlet
boundary values. Also for certain boundary values trapped and bouncing
Bloch fronts coexist.

  In the sharp front and slow velocity regime for the FN model, 
we were able to give a mathematical mechanism for the trapping, bounce 
and coexistence of the two. This involves the coupling of the front velocity
$c$ and front position $x$ close to the boundary. Essentially, Dirichlet
boundary conditions act as a barrier to Bloch fronts coming in. It is
shown that this barrier may or may not be penetrated depending upon 
how fast or far away the incoming Bloch wall is created.
 It is never penetrated, if $v_b<0$. It is always penetrated
if $v_b>1/q^2$.

One could also derive how $c$ and $x$ are coupled close to a boundary for
zero flux boundary conditions, by using a modified Green's function,
with a zero derivative at the boundary.
Zero flux boundary conditions \cite{meron1} show rich behavior too,
(breathing fronts) and the dynamics can be explained by such a calculation.

\end{document}